\documentclass[fleqn,10pt]{wlscirep}
\usepackage[utf8]{inputenc}
\usepackage[T1]{fontenc}
\usepackage{amsmath}

\title{Urbanization, economic development, and income distribution dynamics in India}

\author[1,2,*]{Anand Sahasranaman}
\author[2]{Nishanth Kumar}
\author[3,4]{Lu\'is M. A. Bettencourt}

\affil[1]{Independent Researcher, Chennai, India}
\affil[2]{Centre for Complexity Science, Dept. of Mathematics, Imperial College London, London, SW7 2AZ, UK}
\affil[3]{Dept of Ecology and Evolution, Dept of Sociology, University of Chicago, Chicago IL 60637, USA}
\affil[4]{Santa Fe Institute, Santa Fe NM 87501, USA}
\affil[*]{basking@gmail.com}


\begin{abstract}
India's urbanization is often characterized as particularly challenging and very unequal but systematic empirical analyses, comparable to other nations, have largely been lacking. Here, we characterize India's economic and human development along with changes in its personal income distribution as a function of the nation's growing urbanization. On aggregate, we find that India outperforms most other nations in the growth of various indicators of development with urbanization, including income and human development. These results are due in part to India's present low levels of urbanization but also demonstrate the transformational role of its cities in driving multi-dimensional development. To test these changes at the more local level, we study the income distributions of large Indian cities to find evidence for high positive growth in the lowest decile (poorest) of the population, enabling sharp reductions in poverty over time. We also test the hypothesis that inequality-reducing cities are more attractive destinations for rural migrants. Finally, we use income distributions to characterize changes in poverty rates directly. This shows much lower levels of poverty in urban India and especially in its largest cities. The dynamics of poverty rates during the recent COVID-19 pandemic shows both a high fragility of these improvements during a crisis and their resilience over longer times. Sustaining a long-term dynamic where urbanization continues to be closely associated with human development and poverty reduction is likely India's fastest path to a more prosperous and equitable future. 
\end{abstract}
\begin{document}

\flushbottom
\maketitle

\thispagestyle{empty}

\section*{Introduction}
Cities in India are often seen as extremely problematic and unequal. High rates of poverty, lack of adequate housing and basic services~\cite{Patel_2020}, air pollution~\cite{ravi}, and a profusion of informal settlements (or "slums")~\cite{pandey_infrastructure_2022,sah3} and associated challenging living conditions~\cite{jungari_violence_2022,sah3} are frequently pointed out as strong negative consequences of India's current urbanization. 

None of these challenges should be minimized. However, they have always been associated with low levels of urbanization~\cite{bettencourt_introduction_2021}. Significant improvements in all aspects of development - physical, economic, social - have historically only happened as nations urbanize and gradually acquire the knowledge and capacity to implement systematic policies for infrastructure and service delivery, public health, social support, and greater justice. To put it more bluntly, no nation has ever achieved high levels of development without urbanization~\cite{bettencourt_introduction_2021}. 

The unfolding of these processes systemically and over time remains, however, relatively poorly understood so that development policy is rarely created with an explicit reference to cities. As we write, India remains little urbanized with only $\sim 35$\% of its population living in cities \cite{un}. As a result, despite its large population and accelerating change, India displays the characteristics of relatively underdeveloped cities and nations, including low incomes and high rates of poverty.  

Here, we reconcile various aspects of this apparent contradiction by analyzing India's patterns of development as a function of its urbanization. We perform a multi-scalar analysis, starting with national patterns of change, which we analyze comparatively to other nations. We then analyze the recent dynamics of income distributions in urban versus rural India and in five of its largest cities, with an eye towards better understanding inequality and processes of poverty reduction.  

We choose these two sources of evidence because they allow us to compare India to other nations not purely in time, but at the same level of urbanization.  The positive relationship between national urbanization rates and economic growth is among the most commonly cited patterns relating to urbanization \cite{jones_new_2010,cic1,cic2,lew,gros,bettencourt_introduction_2021}. This effect is the result of a combination of two factors. First, income and most other economic quantities in cities exhibit agglomeration (or network) effects, leading to larger nominal incomes per capita with city size ~\cite{bettencourt_introduction_2021,beal,scot,brau,gla,bair,mum}. Second, at times of fast urbanization there is preferential population migration from less urbanized places and regions to larger cities creating a strong sorting effect~\cite{bettencourt_quantitative_2020,bettencourt_hypothesis_2013,bettencourt_introduction_2021}. As a result, urbanization, especially early in the process, has been associated with reductions in poverty rates \cite{zund_growth_2019,chr,wang,chen,rav1,brelsford_heterogeneity_2017,bettencourt_street_2023}.  We will show that these dynamics are also at play in India.

On the other hand, the relationship between urbanization and income inequality varies across contexts. Income inequality is increasing in some nations, decreasing in others, and exhibiting no effect in yet other cases \cite{brelsford_heterogeneity_2017,sid,wu,ha,sul,adam,garo,haw}. Kuznets (1955) \cite{kuz} theorized an inverted U-shaped relationship between income inequality and economic growth, but this is less characteristic of income than of public goods such as service delivery~\cite{brelsford_heterogeneity_2017}. More recent theory posits that the nature of the relationship is a function of education and skill levels of rural migrants and the staffing capacity of urban firms \cite{sid,jon}. There is therefore a question of whether urbanization in India is increasing or decreasing poverty and inequality, where and at what scales. 

Over the last three decades, urbanization in India has coincided with higher economic growth and declining poverty rates~\cite{cali,dat1,dat2,sah1,sah4,sah5}. 
However, these aggregate dynamics are very heterogeneous across time and different places, with the strongest poverty reductions being urban and relatively recent.  For example, a recent study of long-term dynamics reveals that, even as the overall poverty rate has been declining over time, households at the bottom of the income distribution find an escape from poverty infeasible~\cite{sah1}. In addition, poverty declines in India have also been accompanied by rising income inequality, disproportionately felt by those at the bottom of the income distribution \cite{kan, cha,sah2,sah4,sah5}. 

In this article, we approach these issues empirically by quantifying the relationships between urbanization, economic development, and income distribution in Indian cities. Specifically, we address the following questions: 
\begin{enumerate}
    \item What is the general relationship between national urbanization and economic development?
    \item  What is the relationship between urbanization and the distribution of income in Indian cities?
    \item What is the role of the largest cities in alleviating poverty and economic fragility of households?
\end{enumerate}

Our results support the transformational role of Indian cities in national economic development and poverty reduction, and provide multi-dimensional empirical targets and expectations for its future patterns of urbanization.

\section*{Results}

\subsection*{National urbanization and development}
As nations urbanize, a general structural transformation takes place driving many important development outcomes in tandem~\cite{jones_new_2010,romer_urbanization_nodate,bettencourt_introduction_2021,bettencourt_quantitative_2020}. 
A simple but general way to parameterize these transformations is via an exponential function of the form:
\begin{equation}
\ln y(u) = a + b u  \rightarrow y(u)= A e^{b u}
\label{eq:eq-1}
\end{equation}
where, $y$ is the variable of interest (measured per capita) at the national level, $u$ is the nation's fraction of the population living in cities (in percent), and $a$ and $b$ are constants. The parameter $a$ is the baseline value for a non-urbanized nation - for example, for $y$ as Gross Domestic Product, $A = e^a \simeq 250$ real \$/person/year, much below the the international standard of poverty for lower-middle income nations of \$3.65 per day \cite{wb1}. The parameter $b$ measures the rate of change in $y$ with urbanization, $u$. Specifically, the percent change in $y$ with each percent change in $u$ is given by

\begin{equation}
b = \frac{d \ln y }{d u}.
\label{eq:eq-b}
\end{equation}

\begin{figure}[ht]
\centering
\includegraphics[width=\linewidth]{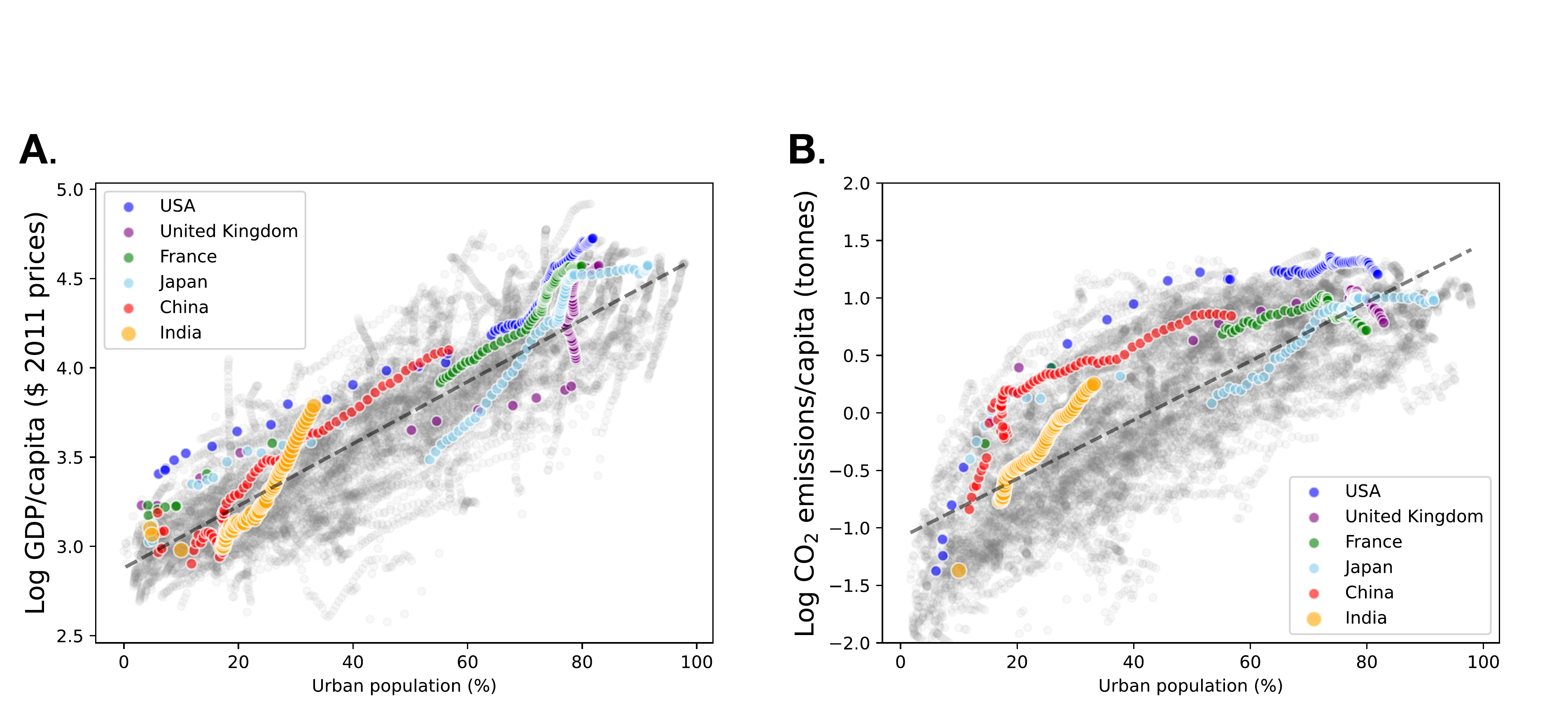}
\caption{Change in urban quantities with urbanization rate across nations and time. A. Change in GDP per capita. B. Change in CO2 emissions per capita}
\label{fig:f0}
\end{figure}

It can be shown that $b$ is not a constant in $u$ over the entire range of urbanization~\cite{bettencourt_quantitative_2020}. This is easy to understand since nations can continue to increase their income per capita (and other indicators) after they are fully urbanized: city states like Singapore or Hong Kong provide examples. However, over most of the range of intermediate levels of urbanization, $b$ can be measured as an approximate constant as seen by the slopes in Fig.~\ref{fig:f0}. We will use the value of $b$ estimated via this approximation to compare urbanization in India to other nations.  

The best known of these relationships takes $y$ as the per capita national income (Gross Domestic Product, GDP$_{\rm pc}$), adjusted for price parity across time and nations (Fig.~\ref{fig:f0}A). This is the most familiar metric of national economic development that is comparable over time and across nations.  The systematic average increase in GDP$_{\rm pc}$ of $b=4-5$\% with each percent increase in urbanization across nations, can be interpreted as strong evidence for the fact that urbanization is a necessary condition underlying national economic growth~\cite{jones_new_2010,zund_growth_2019,brelsford_heterogeneity_2017,bettencourt_street_2023}. Though there are strong variations in context, history and economic structure, all nations with very high GDP$_{\rm pc}$ are highly urbanized. Put another way, there are no predominantly rural rich nations. 

This claim is actually stronger because it is not purely economic: it extends to most other standard measures of development. Table~\ref{tab:table1} shows that exponential increases in development metrics with urbanization, Eq.~\ref{eq:eq-1} is very general and can be used to benchmark nations by their rates of improvement across different quantities. This is a direct way to measure each nation's development returns to urbanization. In doing this, we are not advocating that Eq.\ref{eq:eq-1} is the best possible functional form to fit all the variables in Table~\ref{tab:table1}, we are just using it as a means of national comparison. 

We now use these benchmarks to analyze Indian returns to urbanization, and its multi-faceted process of development in light of other relevant nations. We see from Table~\ref{tab:table1} that many different quantities vary consistently with urbanization rate, and that India is exceptionally successful in its patterns of urbanization in every one of these dimensions. While we discuss caveats below, taking the data at face value shows that Indian GDP$_{\rm pc}$ has increased faster with urbanization than for most other nations, with $b=7.65\%$. In fact, the effect is close to double the average across nations ($b=4\%$), and is faster than similar increases experienced by the US ($b=3.83\%$) and China ($b=5.77\%$), both countries having urbanized earlier, in different eras and very different technological and political contexts. 

Indian urbanization is also very energy intensive - the rates of increase in energy use per capita and CO$_2$ emissions with urbanization are more than twice the global average, and considerably higher than China's. It is important to note that this observation refers to increases in energy use per capita with percent increases in urbanization, but that India still uses very low absolute levels of energy per capita compared to more urbanized and richer nations. This is specifically true when compared to China at the same level of urbanization, see Fig.~\ref{fig:f0}. This measurement does indicate, however, that Indian urbanization and development can (and maybe must) become much less carbon intensive as they continue to increase.  

In terms of other quantities, such as educational attainment and human development, and decreases in child mortality and death rates from infectious diseases, Indian urbanization similarly shows considerably larger returns than the average over nations, and the US and China in particular.  

\begin{table}[ht]
\centering
\begin{tabular}{c|c|c|c|c}
\hline
 {\bf Growth rate, $b$} &  India & USA &  China & All Nations \\
\hline
GDP/capita & 7.65\% (0.96) & 3.83\% (0.24) & 5.77\% (0.35)&  4.00\%  (0.05)\\
Human Development Index & 5.19 (0.14)  & 0.83 (0.08) & 1.38 (0.08)  & 1.02 (0.03) \\ 
Years of Schooling & 15.78\% (2.01) & 1.95\%  (0.14) & 6.23\% (4.68) & 2.86\% (0.14)  \\
Life Expectancy & 3.25 (0.11) & 0.9 (0.07) & 1.18 (0.23) & 0.69 (0.01) \\
Child Mortality Rate &
-11.24 (0.40) & -11.71 (0.91) & -6.44 (0.55) & -3.82 (0.06) \\
Death Rate Infectious Diseases & -13.95\% (0.35) & -12.74\% (1.38) &  -5.65\% (0.19) & -4.34\% (0.13)\\ 
Energy use/capita & 11.86\% (0.36) & -0.60\% (0.55)& 5.60\% (0.48) & 5.63\% (0.10) \\
CO$_2$/capita & 14.04\% (0.55) & 0.06\% (0.62)& 6.73\% (0.84) & 5.89\% (0.09) \\
\hline
\end{tabular}
\caption{Growth rate $b$ of various quantities with each 1 \% increase in urbanization}
\label{tab:table1}
\end{table}

This evidence may seem counterfactual to observers of Indian cities, so we want to clarify the nature of the results. They do not imply that India has currently very high indices of wealth or development, but simply that its increases in urbanization (from a very low level, estimated at $\sim$35\% in 2018 \cite{un}) are associated with very large changes in every dimension of development. All of these quantities started from very low values, as is typical of nations with very incipient urbanization. Thus, these statements emphasize urbanization as a process that - if extrapolated under present trends - would lead potentially to a (very) rich and developed nation, when and if it reaches higher levels of urbanization characteristic of the US (at 85\% urbanization) and other high income nations, and even of China (65\%). 

A few additional caveats on the data are in order. First, there is a general expectation that India's level of urbanization ($u$) is likely being underestimated by official statistics \cite{sri,denis,onda}. A substantial underestimation could explain in part the over-performance across every dimension of development. For example, while the current level of urbanization in India is $u\simeq 35\%$ \cite{un}, this estimate follows from a conservative definition of urban areas by national statistics. Other estimates, have suggested larger $u$, such as 43\% in 2011 \cite{onda}. 

We can contextualize the meaning of these different urbanization rates on urban quantities such as income and human development, using Eq.~\ref{eq:eq-1} and the data in Table~\ref{tab:table1}. Let us suppose that the actual urbanization rate $u^*$ is higher than estimated ($u$), such that $u^* = \lambda u$,  with $\lambda>1$.   We can then ask what the estimated level of urbanization in India would have to be, to match the growth rate of GDP$_{\rm pc}$ on average across nations. For example, we can compute $\lambda$, such that the return on per capita GDP is the same for India ($b^*$) as the average across nations ($\bar b=4\%$):

\begin{eqnarray}
\ln y = a + b u = a + \frac{b}{\lambda} u^*  \rightarrow  \lambda = \frac{b^*} {\bar b}. 
\end{eqnarray}

\noindent This results in $\lambda = 1.91$, and, consequently, India's urbanization rate being $u^* = 59\%$. This would mean that the actual urbanization rate (in 2011), would have been close to 20\% higher than what the definition of ``urban" in India would lead us to believe, which seems implausibly high.

However, if we assume that the actual urbanization rate was indeed higher, not as much as 59\%, but say at 43\%, as estimated in\cite{onda}, then what would be the implications for growth rate of per capita GDP with urbanization? In this case, $\lambda=1.39$. This would result in the growth rate of GDP$_{\rm pc}$ $b^*=5.5\%$, which is close to the estimate for China. 

Despite these qualifications and uncertainties, these findings suggest that urbanization in India is associated with an extremely successful, multi-dimensional process of development, at least on average across its vast and diverse population. To address the question of variations within this population, we now investigate
the dynamics of growth across the income distribution. This allows us a deeper understanding of poverty and distributional dynamics across India, and specifically, in its largest cities.   

\subsection*{Distributional dynamics of personal income in India}

We now ask if the increases in national income per capita observed in the aggregate are benefiting all of India's population equally. As suggested by the previous section, and underlying findings in urban science \cite{sah6,sah7}, urban India performs quite differently. Surprisingly, perhaps we will also find that different cities in India have experienced different types of income dynamics, helping us understand contextual factors and leaving scope for policy. 

To set the stage, studies of long-term changes in the Indian income distribution have shown that the income share of the bottom half of the distribution increased for the first three decades after independence. However, since the 1980s this share has seen a steep decline, dropping from 24\% in 1981 to 15\% in 2015. During the same time period, the share of the top 10\% increased from 31\% to 56\%  \cite{cha}. More recent work \cite{sah5} shows that, between 2014-19 there has been an increase in income share in the middle of the income distribution (middle class), but the very bottom of the income distribution has seen a decline, with share of the bottom decile falling from 1.6\% to 1.2\%. Analysis of the Growth Incidence Curve (GIC) reveals that real income growth rates are positive (5-6\% per annum) over the bulk of the income distribution, except for the bottom ventile, which experienced real income decline of -5\% per annum \cite{sah5}. 

Given this backdrop, we now seek to delve deeper into the dynamics of India's income distribution, and focus on its urban component and its largest cities. We construct income distribution for different geographies using data for the period 2015-2019, from the Consumer Pyramids Household Survey (CPHS) published by the Centre for Monitoring Indian Economy \cite{vyas1} (details on construction of income distributions in Materials and Methods section). There is concern that the poorest households are under-represented in the CPHS data, and as with any survey of this kind, the richest are unlikely to participate \cite{jha,drez,vyas2}. Comparing labour income estimates from CPHS data with the Periodic Labour Force Survey (PLFS) by the Government of India reveals that CPHS mean income estimates are higher than PLFS estimates for the all-India distribution, but that the mean income estimates are almost identical for the urban income distribution \cite{jha}. Both surveys are also found to concur on certain stylized facts about the Indian workforce \cite{jha}. 

We construct the income distributions separately for urban and rural India, and for each of the individual cities under consideration. This means that the levels of income in each decile (or percentile) will be different for income distributions of different cities. Noting these differences allows us to draw conclusions about the relative economic well-being of populations located at the same position in different income distributions.

We start by first focusing on rural-urban differences. Prior work has revealed that when the Indian income distribution is decomposed into a rural and an urban components, the decline in real income observed between 2014-19 is confined only to the bottom decile of the {\it rural} distribution (-5\%) \cite{sah5}. The bottom decile of the urban distribution was found to experience real income growth of 5.6\% \cite{sah5}.  

\begin{figure}[ht]
\centering
\includegraphics[width=\linewidth]{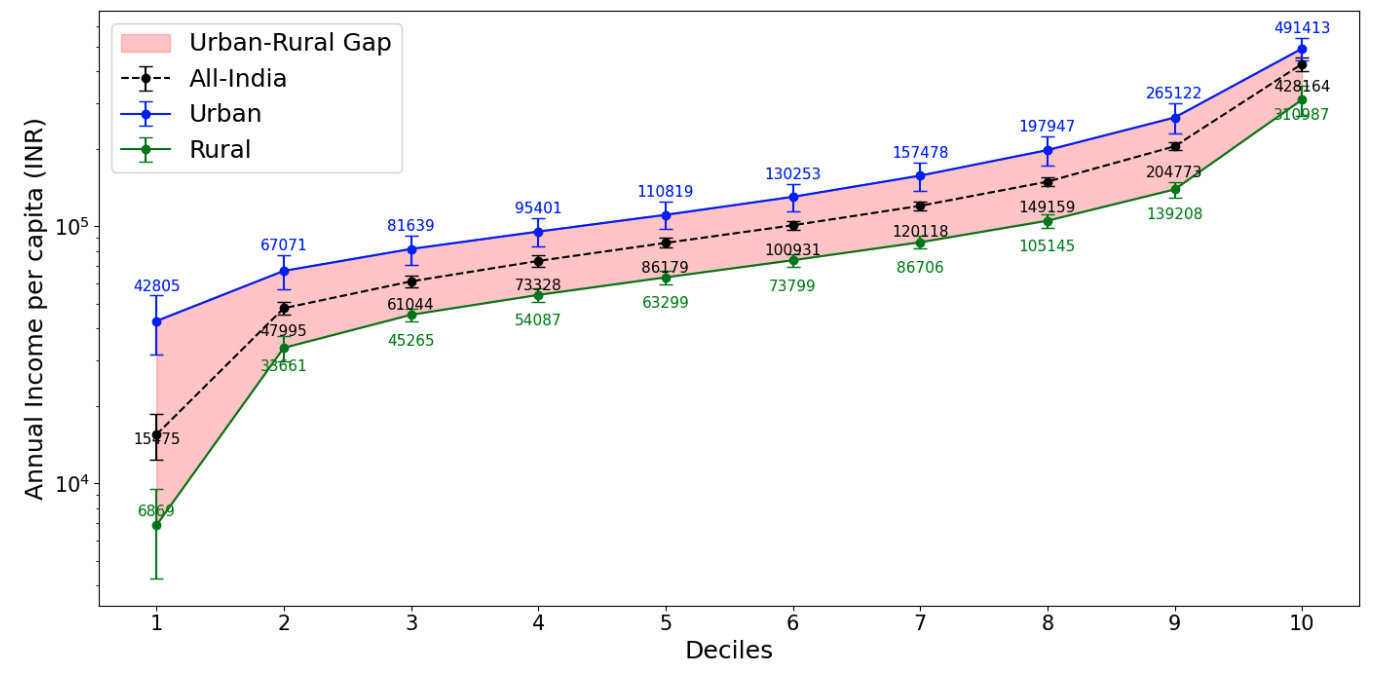}

\caption{Income levels in rural and urban India across deciles (2019). Average per capita income levels of deciles of the urban, rural, and all-India income distributions in 2019. Urban incomes are always higher than rural incomes across all deciles. Errors bars show standard errors. }
\label{fig:f2}
\end{figure}

Given this context, we study the average per capita (nominal) income for the bottom 5 deciles in India's urban and rural income distributions. We find that, in addition to urban income growth rates being substantially higher than rural growth rates at low levels of income, the absolute levels of average per capita income are also substantially higher (Fig.~\ref{fig:f2}). This is particularly pronounced in the bottom decile, where the average urban income is over 5.5 times the rural income. We also find, for instance, that the median per capita income (50\textsuperscript{th} percentile) in the rural Indian income distribution (in 2019) is INR 80,102 (USD 1 $\approx$ INR 80), which corresponds to the 23\textsuperscript{rd} percentile in the urban income distribution. Cities, in this specific sense, provide a significant income premium over villages for their poorest citizens.

From the overall urban income distribution, we can now zoom in into the income distributions of five large metropolitan areas - Delhi, Mumbai, Kolkata, Chennai, and Hyderabad. Despite the fact that the CPHS survey does not claim to be representative at the level of these cities, it captures a wide swathe of the income distribution, which makes it a valuable data source for the present analysis. We only focus on the largest cities because the survey data available to construct the income distribution becomes increasingly sparse for smaller cities. 

We compute the Growth Incidence Curves (GICs), which capture the extent of real income growth occurring in different parts of the distribution, for these five cities from 2015-19. This gives us a sense of how overall income growth has been distributed across different income classes in society. For details on computation of GICs and real income growth, see Materials and Methods.

\begin{figure}[ht]
\centering
\includegraphics[width=\linewidth]{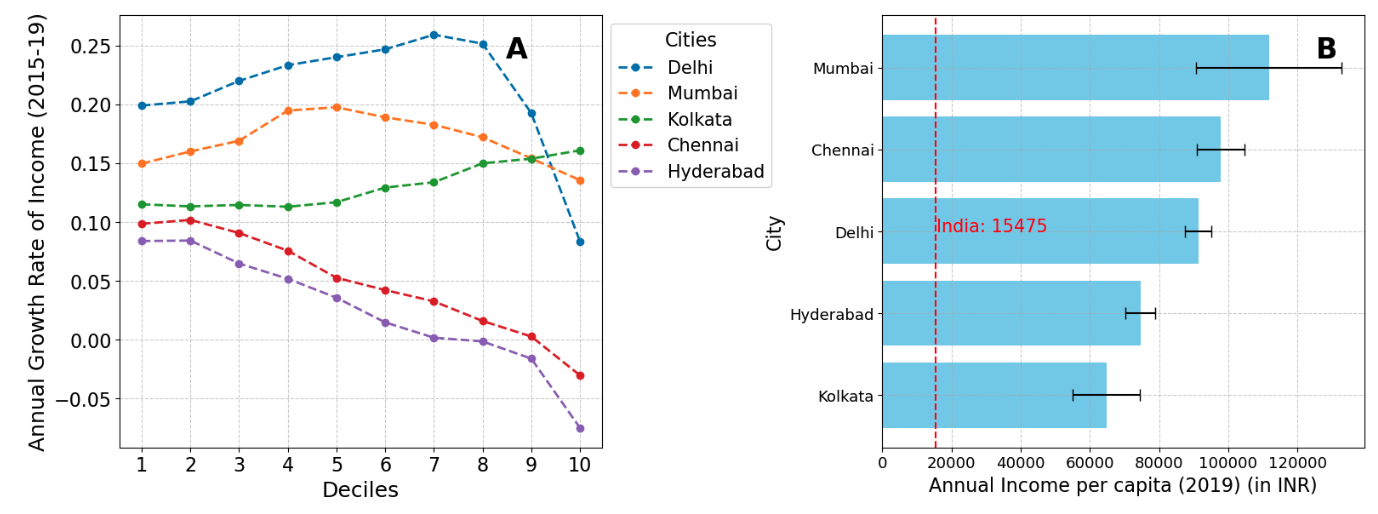}
\caption{Income distribution dynamics in large Indian cities. A. Decile-wise Growth Incidence Curves for five large cities of India (2015-19). The poorest decile shows strong positive income growth in all cases. B. Per capita Income level (INR) for the bottom decile (2019). Average income levels at the bottom of the income distribution for these cities are significantly higher than those of the overall Indian income distribution (dashed line: INR 15,475). Black bars indicate standard error.}
\label{fig:f4}
\end{figure}

We find that real income growth is strongly positive for the bottom decile of the income distribution (the poorest), with annual growth rate across all cities greater than 8\% (Fig.~\ref{fig:f4}A). The improved economic prospects of the poorest sectors of the population in large Indian cities are confirmed when we study the average per capita income of the bottom decile in these cities - they are 4 to 7.5 times higher than the per capita income of the bottom decile of the overall Indian income distribution (Fig.~\ref{fig:f4}B). Additionally, the 50\textsuperscript{th} percentile in the rural Indian income distribution (which corresponds to the 23\textsuperscript{rd} percentile in the urban income distribution), also corresponds to the following: the 13\textsuperscript{th} percentile of Kolkata's income distribution, 7\textsuperscript{th} of Hyderabad, 3\textsuperscript{rd} of Delhi, and 2\textsuperscript{nd} of Chennai and Mumbai. This analysis indicates that while cities, in general, provide opportunities for improved economic prospects, the effect is particularly pronounced in the largest cities. 

We also find that real annual growth rates are rising as we move higher in the income distribution for Mumbai, Delhi and Kolkata, indicating likely increased income inequality in these cities. By contrast, income growth rates decline as we move higher in the income distribution for Chennai and Hyderabad, indicating reductions in income inequality.  

Given these differences, we consider whether population growth trends in cities that are inequality-reducing are different from the other cities. While we do not have official population statistics post 2011 due to the postponement of the Indian Census, there is a recent study from the International Institute for Population Sciences which estimates population growth rates for urban agglomerations in India \cite{dey}. The study estimates that the populations of Hyderabad and Chennai grow at the fastest rate between 2011 and 2021, with population growth of 11.1\% and 8.8\% between 2015-19, respectively. Comparatively, Mumbai and Kolkata showed growth rates of 4.9\% and 3.6\%, respectively. While this evidence is circumstantial, it suggests the (rational choice) hypothesis that inequality-reducing cities are more attractive to poorer migrants moving out of rural areas to improve their economic prospects. 

The absence of the highest incomes from the survey data prevents us from making robust estimates of income inequality in the cities under study. However, multiple studies suggest that income inequality in India, as a whole, has been growing over the last two decades \cite{kan,cha,sah2,sah4,sah5}. As has been observed in other countries \cite{sark,arv,cast,roy}, it is likely that the extent of inequality is higher in the largest urban centres due to the prevalence of people with the very highest incomes. The rise of the billionaire class in India is an exclusively urban (primarily large city) phenomenon, and as accounted by the Forbes billionaire list, the number of billionaires has gone up from 9 in 2000 to 200 in 2024 \cite{fbs}. 

Given the dynamics at the bottom of the income distributions in cities, we now turn our attention to their impact on urban poverty trends. 

\subsection*{Poverty levels and economic fragility}
We now complement our analysis of income dynamics, with an explicit focus on poverty reduction. Increasing urbanization has, in general, been found to be associated with declining poverty levels in nations around the world~\cite{ngu,chr,wang,chen,dat1,dat2}. India's poverty rate has historically remained stubbornly high, at over 50\% from independence in 1947 until the late 1970's. However, it has since declined substantially from 53\% in 1978 to 21\% in 2006 \cite{dat2}. Since 1991, several studies have found urban growth to be the primary driver of both rural and urban poverty reductions \cite{dat1,dat2}, in agreement of our findings above.

To analyze poverty dynamics in urban India, we adopt the World Bank's definition of the poverty line for lower-middle income countries. This is the proportion of population with incomes below \$3.65 per day, based on Purchasing Power Parity (PPP). This adjustment for cost of living allows for global comparability of income and consumption across nations. For a description of how we translate CPHS nominal income data into PPP terms, see Materials and Methods.

\begin{figure}[ht]
\centering
\includegraphics[width=\linewidth]{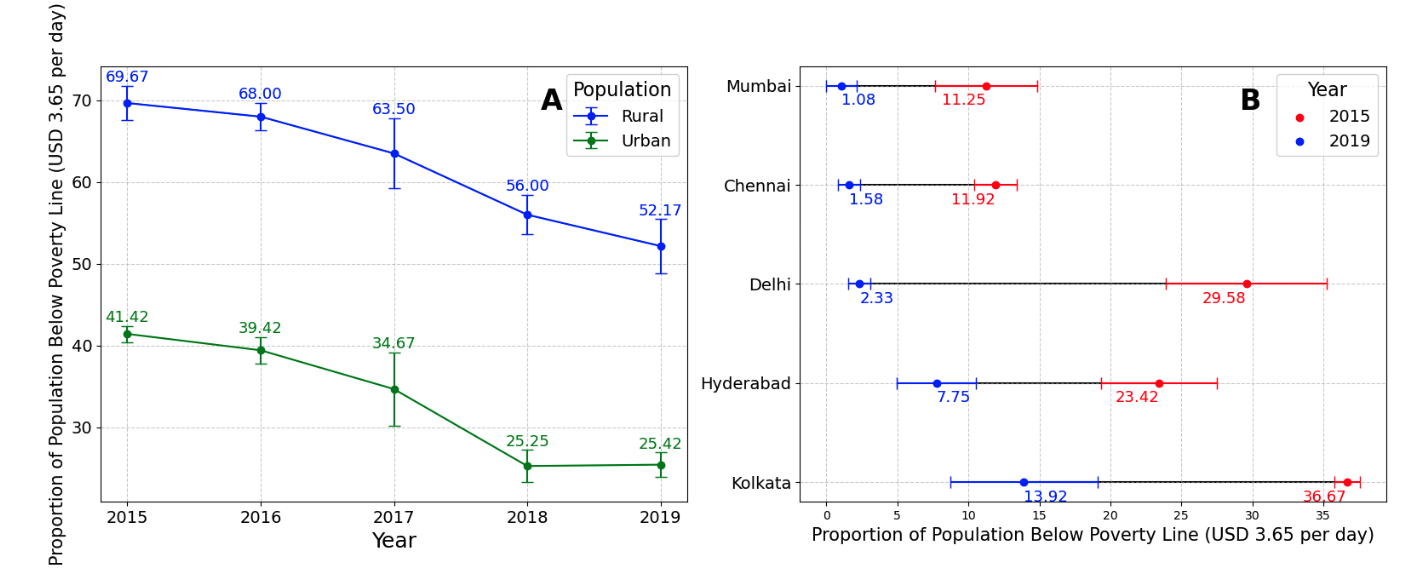}
\caption{Poverty rate in urban India. A. Proportion of population below USD 3.65/ day PPP in urban and rural income distributions (2015-19). Poverty declines in both rural and urban India, but urban India has a much lower poverty rate. B. Proportion of population in poverty in five Indian cities. Poverty levels show a decline between 2015 and 2019 across all cities. Black bars indicate standard error.}
\label{fig:fpov1}
\end{figure}

We find that poverty rates in both rural and urban India declined between 2015-19 (Fig.~\ref{fig:fpov1}A). The decline in rural poverty, despite lacklustre agricultural growth, points to urban economic growth driving rural poverty trends, in agreement with the findings of Datt, Ravallion, and Murgai~\cite{dat1}. In addition, we also find that urban poverty is significantly lower than rural poverty, in line with expectations.

The role of cities, especially large ones, in poverty reduction in India is brought into sharp relief when we focus on the poverty levels observed in the five cities in our analysis. We find that 2-4\% of the population of Delhi, Mumbai, and Chennai, and 8-14\% of Hyderabad, and Kolkata would be considered poor by this definition in 2019 (Fig.~\ref{fig:fpov1}B). This proportion is significantly smaller than 11-37\% of population that was poor in these cities in 2014. Importantly, the proportions of the poor in the largest cities are also significantly lower than the overall poverty level of urban India, which we compute to be 25\% in 2019. 

It is important to point out that we use a constant poverty line of \$3.65 per day for both rural and urban India, despite local cost of living differences. Using per capita consumption expenditure data from the Household Consumption Expenditure Survey (2022-23)\cite{nss2}, we find that the ratio of urban to rural consumption expenditure is 1.70, indicating that, on average, cost of living in cities is 1.7 times higher than in rural areas. If we scale up the urban poverty line by a factor of 1.7 to account for the cost of living difference, we find that across the 5 cities in our analysis, the average fraction of population below this adjusted poverty line drops from 54\% in 2015 to 34\% in 2019. This is still lower than the rural poverty rate of 52\% in 2019, and indicates that the poverty alleviating effects of these cities are not just an artefact of cost of living differences.

The dynamics of economic activity in larger cities yield super-linear scaling of average income in India \cite{sah6,bettencourt_hypothesis_2013,bettencourt_introduction_2021}, and our findings here indicate that this scaling behaviour could extend all the way down the income distribution to the poorest populations, enabling greater poverty reduction in the largest cities. This result therefore supports the critical role of the largest cities in the Indian urban hierarchy in combating poverty.  

\begin{figure}[ht]
\centering
\includegraphics[width=\linewidth]{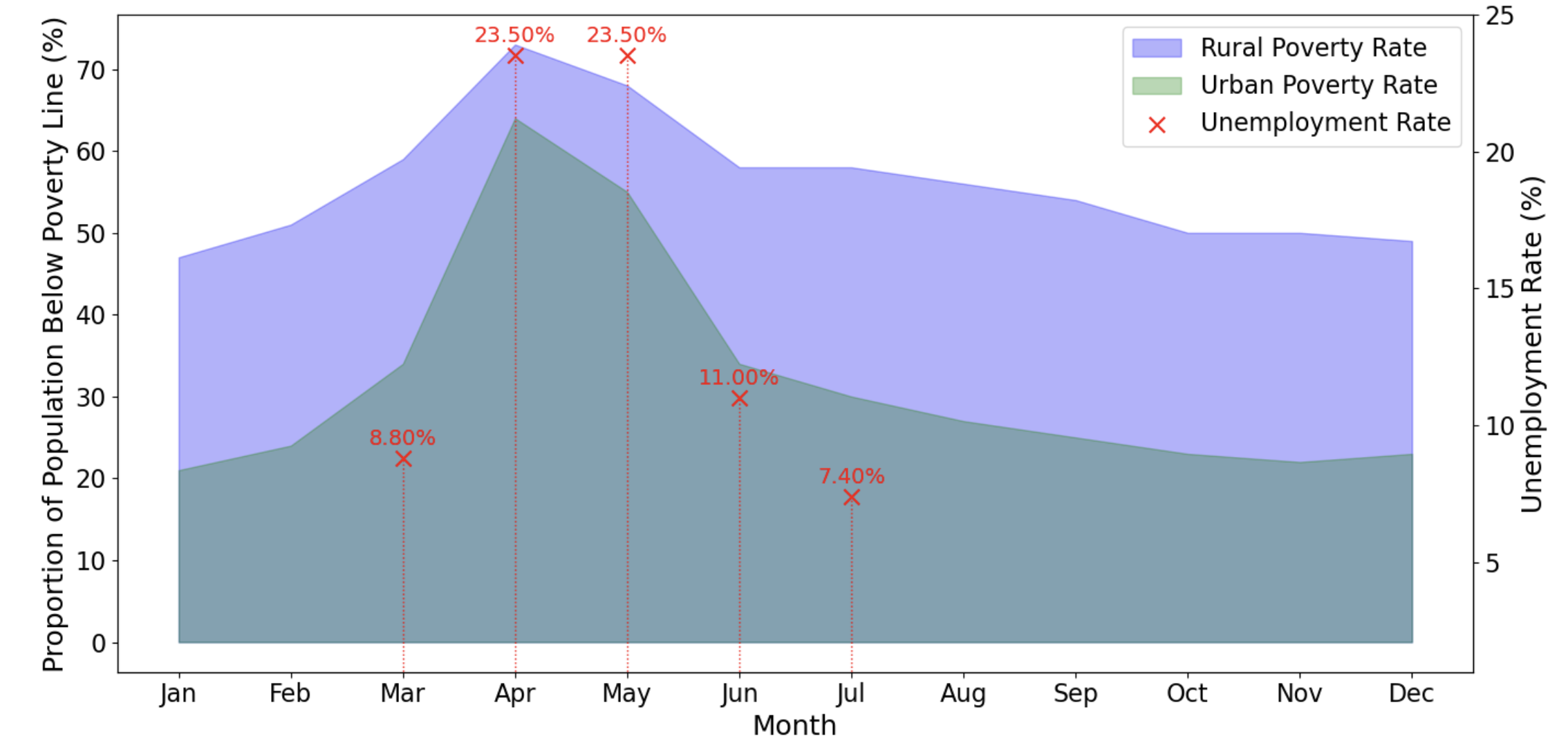}
\caption{Poverty in 2020. Proportion of population below \$ 3.65 per day in urban and rural income distributions. Trends in unemployment rate indicate that poverty rate closely mirror unemployment. Black bars indicate standard error.}
\label{fig:fpov2}
\end{figure}

Finally, we seek to investigate the economic fragility (or robustness) of urban households around the poverty line. In prior work, we have shown that even as overall poverty levels decline in India, there are many households cycling in and out of poverty over time~\cite{sah1}. The COVID-19 pandemic presents us with a natural experiment to observe the impact of an exogenous shock on the economic resilience of households around the poverty line. We study the monthly proportions of population earning below \$ 3.65 per day in 2020, for the urban and rural income distributions. We find that although there is a spike in both rural and urban poverty upon the onset of COVID-19 and the announcement of a nationwide lockdown in March 2020, the urban poverty spike is much more pronounced (Fig.~\ref{fig:fpov2}). 

Over the subsequent two months, unemployment rose rapidly to 23.5\%, coinciding with a sharp increase in the proportion of population in poverty, peaking at 73\% in rural India (from 47\% in January 2020) and 64\% in urban India (from 21\% in January 2020). These stark effects last for about 3 months. Starting June 2020, as employment starts picking up again, urban poverty levels trend sharply downwards, such that by the end of 2020, urban poverty levels are just about 2\% higher than before the pandemic.  

These fast movements in poverty upon the onset of COVID-19 are indicative of the fragility of incomes in the lower half of the distribution, but also illustrate the dynamical robustness of urban environments in poverty reduction.

\section*{Discussion}
We developed a comparative analysis to measure the effects of urbanization on various national development outcome metrics. We showed that India's current urbanization outperforms average national expectations for development from across the world, including in China and the US. We discussed to what extent this observed outperformance may be the result of overly restrictive urban definitions in Indian national statistics (Census of India). Analyzing a range of higher estimates, we conclude that it is still likely that India's urbanization is rather successful, yielding faster multi-dimensional development than most other nations, with each additional percent of urbanization. It remains an interesting question to what extent this outperformance can be sustained as India continues to urbanize from the current low levels of different development indicators.

We substantiated and refined these aggregate findings by analyzing survey data on income distribution dynamics comparing urban with rural India and to five of its largest cities. We showed that the poorest decile of the Indian income distribution is primarily composed of rural incomes, and that urban centers are associated with significantly higher absolute income levels~\cite{sah5}. Income distributions for the largest Indian cities reveal robust positive income growth even at the bottom of the distribution.  This income growth is occurring in a context where the largest cities attract the most migrants and provide higher levels of basic infrastructure and services, compared to smaller cities and villages \cite{sah3}. Even amongst the larger cities, we posit the hypothesis that inequality reducing cities (those with progressive GICs) are more attractive to poorer migrants, and are likely to see higher population growth - a trend that seems to be borne out in our small sample of five cities.

The dynamics at the bottom of the urban income distribution are in sharp contrast to the much lower income levels and growth rates in the rural income distribution, and thus provide an impetus for increased urbanization, especially to the larger cities. Income inequality, on the other hand, is likely to be higher in urban India and trending higher over the last two decades, with larger cities being more unequal than smaller cities and rural areas. This is line with concerns around the world on rising economic inequality and the increasing condensation of income and wealth among those at the very top of these distributions.

Poverty in India is much more concentrated in rural areas than in cities, with deep poverty an exclusively rural phenomenon~\cite{sah4,sah5}. Studying the extent of poverty in large Indian metropolises reveals that these cities, on average, have even lower levels of poverty that urban India as a whole. We also find that the strong, positive growth rate of income in the bottom decile of these cities is reflected in sharply reducing poverty trends over time. Once again, this result highlights the central role of cities atop India's urban hierarchy in poverty reduction and economic growth. An important caveat to this reduction in poverty is the fragility to shocks at the bottom of the income distribution. Analysis of the income distribution during the intensive phase of the COVID-19 pandemic shows that the population in the lower half of the distribution lacks short-term income resilience, though it recovers on longer time scales. A much more detailed study, including a better understanding of savings and public and private social supports, is necessary to understand the dynamics of income and poverty reduction in India.

Overall, the picture that emerges from our analysis is rather positive for Indian urbanization. We want to make clear that this is not an endorsement of the current state of Indian cities and their strong economic inequalities, their lack of basic services for large fractions of their populations, or the difficult living conditions in their informal settlements. Rather, analysis of available evidence points out that improvements in many of these issues are being achieved rapidly as the Indian population becomes more urban and that, in this respect, India is doing very well compared to most other nations at comparable (low) levels of urbanization. Though these prospects appear bright, the challenge of continuing to support fast multi-dimensional development in a nation as large and unequal as India - and in some of the world's largest and most complex cities - will remain a major challenge for science and policy in the decades ahead.    

\subsection*{Materials and Methods}
\subsubsection*{Data Sources:}
Data for real per capita national GDP, Human Development Index, mean years of schooling, mean life expectancy, child mortality rate, death rate from infectious diseases, energy use per capita and CO$_2$ emissions per capita are available from the World Bank Development Indicators Databank \href{https://databank.worldbank.org/source/world-development-indicators}{https://databank.worldbank.org/source/world-development-indicators} and from “Our World in Data” \href{https://ourworldindata.org/urbanization}{https://ourworldindata.org/urbanization} . Data to construct the income distributions for India, rural and urban India, and for the five cities (Delhi, Mumbai, Kolkata, Hyderabad, and Kolkata) was obtained from the Consumer Pyramids Household Survey (CPHS), conducted by the Centre for Monitoring the Indian Economy (CMIE). Unemployment data was also obtained from the CPHS survey.

\subsubsection*{Analysis of national urbanization effects:}
As indicated in the main text, we use data on an aggregated economic indicator per capita at the national level $y$. For each year, we take the logarithm of this indicator, creating a time series. This time series is correlated to the time series of urbanization for the same nation, across all nations. The universe of nations and temporal extents depends on the focal indicator $y$. 

The coefficients $a$ and $b$ are then estimated from these time series using standard linear correlation, via least squares minimization.

\subsubsection*{Constructing income distributions:}
The Consumer Pyramids Household Survey (CPHS)  is a quarterly, pan-India, panel survey of 170,000 households, capturing household-level data since 2014, including nominal income data. The dataset is made geographically representative by sampling one or more regions from a state, such that each region is comprised of a set of neighbouring districts with similar characteristics such as agro-climatic conditions, urbanization, and female literacy, as per data from to Census 2011. Each household in the panel is visited thrice a year and all data is captured monthly.

Monthly per-capita income is computed by adjusting the total household income with the reported size of households, using a square root equivalence scale \cite{deat}. The income distribution for each of the cities in our analysis is constructed by weighting each income by an adjustment factor (provided by CPHS), so as to correctly represent different household typologies in the distribution. These adjusted incomes are then added over each percentile, to construct the percentile income distribution for the city. We divide this distribution into $n=10$ equi-sized bins to construct deciles of the income distribution.

\subsubsection*{Computing real income growth and constructing Growth Incidence Curves (GICs):}
In order to compute the cumulative annual growth rate ($CAGR$) of average income ($GDP_{pc}$) between 2015 and 2019, we have:

\begin{equation}
CAGR = {\frac{GDP_{pc}(2019)}{GDP_{pc}(2015)}}^{1/4} - 1 
\label{eq:eq-cagr}
\end{equation}

The CPHS survey captures nominal incomes, and thus the CAGR we compute is the nominal growth rate of income. To translate this into real growth rate ($r$), we subtract the average annual inflation rate ($i$) from the $CAGR$. 

\begin{equation}
r = CAGR - i
\label{eq:eq-r}
\end{equation}

The average annual inflation rate for 2015-19 is 4.14\%, computed as the average of the yearly inflation rates for India for each of the years from 2015 to 2019 (4.9\%, 4.9\%, 3.3\%, 3.9\%, and 3.7\% - data from the World Bank \cite{wb2}).  

In order to construct the Growth Incidence Curve (GIC) for a city, we compute the real growth rate ($r$) at each decile of the income distribution (between 2015 and 2019). The GIC is this sequence of real growth rates along the entire distribution (10 deciles).

\subsubsection*{Computing poverty rates based on \$ 3.65 per day poverty line:} 
To compute the poverty rate for a city corresponding to World Bank's \$ 3.65 per day poverty line, we first determine the proportion of population in India living on less than \$ 3.65 per day, for each of the years from 2015 to 2019 (61\%, 60\%, 55\%, 47\%, and 44\% - from World Bank data) \cite{wb3}. For each year, we then determine a PPP conversion factor that when multiplied to 3.65 yields an adjusted level of income such that the proportion of the population living below this level of income (in the all-India income distribution) matches the observed level from World Bank's poverty data. We compute this adjusted level of income for each year, and this forms the equivalent poverty line in our analysis for that year.  We use these annual poverty levels (for each year between 2015 and 2019) to compute the poverty rates in all the income distributions under study: urban and rural India, Mumbai, Delhi, Kolkata, Chennai, and Hyderabad.

\bibliography{sample}

\section*{Author contributions statement}
AS and LMAB conceptualized the study, contributed methods, analyzed the data, wrote and edited the paper. NK organized the data, contributed methods, analyzed the data, and edited the paper.

\section*{Competing interests:} The authors declare no competing interests 

\end{document}